\documentclass[twoside,preprintnumbers,amsmath,amssymb,showpacs,nofootinbib]{revtex4}
\usepackage{amsmath,amssymb,url}
\usepackage{graphicx,subfigure}

\usepackage{braket,euscript,dsfont}
\usepackage{fancyvrb}

\renewcommand{\d}{\ensuremath{\mathrm{d}}}

\newcommand{\MSbar}{\overline{\mbox{MS}}}
\renewcommand{\d}{\ensuremath{\mathrm{d}}}

\newcommand{\p}{\partial}

\begin{document}

\title{{\bf More on the non-perturbative Gribov-Zwanziger quantization of linear covariant gauges}}
\author{M.~A.~L.~Capri}
\email{caprimarcio@gmail.com}
\affiliation{Departamento de F\'{\i}sica Te\'{o}rica,
Instituto de F\'{\i}sica, UERJ - Universidade do Estado do Rio de Janeiro, Rua S\~ao Francisco Xavier 524, 20550-013, Maracan\~a, Rio de Janeiro, Brasil}\author{D.~Dudal}
\email{david.dudal@kuleuven-kulak.be}
\affiliation{KU Leuven Campus Kortrijk - KULAK, Department of Physics, Etienne Sabbelaan 53, 8500 Kortrijk, Belgium}
\affiliation{Ghent University, Department of Physics and Astronomy, Krijgslaan 281-S9, 9000 Gent, Belgium}
\author{D.~Fiorentini}
\email{diegofiorentinia@gmail.com}
\affiliation{Departamento de F\'{\i}sica Te\'{o}rica,
Instituto de F\'{\i}sica, UERJ - Universidade do Estado do Rio de Janeiro, Rua S\~ao Francisco Xavier 524, 20550-013, Maracan\~a, Rio de Janeiro, Brasil}\author{M.~S.~Guimaraes}
\email{msguimaraes@uerj.br}
\affiliation{Departamento de F\'{\i}sica Te\'{o}rica,
Instituto de F\'{\i}sica, UERJ - Universidade do Estado do Rio de Janeiro, Rua S\~ao Francisco Xavier 524, 20550-013, Maracan\~a, Rio de Janeiro, Brasil}\author{I.~F.~Justo}
\email{igorfjusto@gmail.com}
\affiliation{Departamento de F\'{\i}sica Te\'{o}rica,
Instituto de F\'{\i}sica, UERJ - Universidade do Estado do Rio de Janeiro, Rua S\~ao Francisco Xavier 524, 20550-013, Maracan\~a, Rio de Janeiro, Brasil}\affiliation{Ghent University, Department of Physics and Astronomy, Krijgslaan 281-S9, 9000 Gent, Belgium}
\author{A.~D.~Pereira}
\email{aduarte@if.uff.br}
\affiliation{UFF $-$ Universidade Federal Fluminense, Instituto de F\'{\i}sica, Campus da Praia Vermelha, Avenida General Milton Tavares de Souza s/n, 24210-346, Niter\'oi, RJ, Brasil.}
\affiliation{SISSA $-$ International School for Advanced Studies, Via Bonomea 265, 34136 Trieste, Italy}
\author{B.~W.~Mintz}
\email{bruno.mintz.uerj@gmail.com}
\affiliation{Departamento de F\'{\i}sica Te\'{o}rica,
Instituto de F\'{\i}sica, UERJ - Universidade do Estado do Rio de Janeiro, Rua S\~ao Francisco Xavier 524, 20550-013, Maracan\~a, Rio de Janeiro, Brasil}\author{L.~F.~Palhares}
\email{leticiapalhares@gmail.com}
\affiliation{Departamento de F\'{\i}sica Te\'{o}rica,
Instituto de F\'{\i}sica, UERJ - Universidade do Estado do Rio de Janeiro, Rua S\~ao Francisco Xavier 524, 20550-013, Maracan\~a, Rio de Janeiro, Brasil}\author{R.~F.~Sobreiro}
\email{sobreiro@if.uff.br}
\affiliation{UFF $-$ Universidade Federal Fluminense, Instituto de F\'{\i}sica, Campus da Praia Vermelha, Avenida General Milton Tavares de Souza s/n, 24210-346, Niter\'oi, RJ, Brasil.}
\author{S.~P.~Sorella}
\email{silvio.sorella@gmail.com}
\affiliation{Departamento de F\'{\i}sica Te\'{o}rica,
Instituto de F\'{\i}sica, UERJ - Universidade do Estado do Rio de Janeiro, Rua S\~ao Francisco Xavier 524, 20550-013, Maracan\~a, Rio de Janeiro, Brasil}

\begin{abstract}
In this paper, we discuss the gluon propagator in the linear covariant gauges in $D=2,3,4$ Euclidean dimensions. Non-perturbative effects are taken into account via the so-called Refined Gribov-Zwanziger framework. We point out that, as in the Landau and maximal Abelian gauges, for $D=3,4$, the gluon propagator displays a massive (decoupling) behaviour, while for $D=2$, a scaling one emerges. All results are discussed in a setup that respects the Becchi-Rouet-Stora-Tyutin (BRST) symmetry, through a recently introduced non-perturbative BRST transformation. We also propose a minimizing functional that could be used to construct a lattice version of our non-perturbative definition of the linear covariant gauge.
\end{abstract}
\maketitle

\section{Introduction} \label{Intro}
In the last decade, a great advance in the understanding of the non-perturbative behaviour of the elementary quark, gluon and ghost excitations of gauge-fixed Yang-Mills theory was achieved. Although being a gauge-dependent quantity, the gluon propagator could in principle contribute to our comprehension of e.g.~color confinement and as such it has attracted attention from a very diverse community. Many non-trivial features of the non-perturbative regime of Yang-Mills theories were captured from gauge-fixed lattice results and functional methods for Green functions. A particular approach which was able to produce non-perturbative results is the so-called (Refined) Gribov-Zwanziger framework. In this paper, we will focus on the latter non-perturbative quantization approach.

The Gribov-Zwanziger framework  emerges from a direct attempt of eliminating the well-known Gribov copies, whose existence were pointed out by Gribov in his seminal paper \cite{Gribov:1977wm}. After the usual gauge fixing through the Faddeev-Popov method, a residual gauge symmetry remains and is possible to find gauge fields which are connected via a gauge transformation and satisfy the same gauge condition. Let us refer to the reviews \cite{Sobreiro:2005ec,Vandersickel:2012tz} or the recent paper \cite{Maas:2015nva} (or also \cite{Mehta:2014jla}), where examples of Gribov copies are explicitly worked out. We emphasize that the existence of the Gribov phenomenon is not a pathology of a specific gauge but a general feature of all Lorentz covariant gauges \cite{Singer:1978dk}.

To perform a consistent path integration, that is one in the sense of being free of all local gauge redundancy, the elimination of these copies is essential. In \cite{Gribov:1977wm}, Gribov proposed a partial solution to the problem which consists of restricting the path integral to a region, called the Gribov region, which is free of a large set of copies, namely, infinitesimal ones\footnote{An alternative approach consists of sampling over Gribov copies, worked out in \cite{Serreau:2012cg,Serreau:2015yna} with applications discussed in \cite{Tissier:2010ts,Reinosa:2014ooa,Reinosa:2015gxn}.}. The proposal was worked out just up to first-order in perturbation theory and was generalized to all orders by Zwanziger, \cite{Zwanziger:1989mf}. The result is a local, renormalizable effective action, the Gribov-Zwanziger action, which implements the restriction of the path integral domain to the Gribov region \cite{Zwanziger:1989mf}. This construction was done in the Landau gauge, and explores particular features of this gauge. Therefore, extending this formalism to other gauges, albeit very important, is not trivial. A recent generalization to the $SU(2)$ maximal Abelian gauge was presented in \cite{Gongyo:2013rua}, consistent with earlier proposals of some of us in \cite{Capri:2008vk,Capri:2010an,Capri:2013vka}.

A property of the Gribov-Zwanziger action is that it breaks the standard BRST symmetry \cite{Vandersickel:2012tz}. Nevertheless, the breaking is soft in the sense that it becomes irrelevant when the UV limit is taken. This is due to the appearance of the Gribov parameter, a dynamical mass scale which naturally emerges due to the restriction of the path integral to the Gribov region. In this way, the Gribov-Zwanziger  framework seems to face two challenges. Firstly, its construction is highly gauge-dependent in the sense that particular properties of the chosen gauge must be taken into account, which makes a general construction valid for an arbitrary gauge choice very complicated. Secondly, it lacks BRST invariance, a tool which plays a fundamental role in the proof of e.g.~gauge independence of physical quantities.

Very recently, however, efforts in the opposite direction of such mentioned problems were presented \cite{Capri:2015ixa}. In particular, it was shown that the Gribov-Zwanziger action in the Landau gauge enjoys a modified non-perturbative BRST symmetry, which captures non-perturbative effects of the theory. So, in this way, the usual BRST transformation corresponds to a symmetry at the perturbative level and, when we pass to the non-perturbative regime, a suitable modification of these perturbative transformations is required. In particular, a fully localized non-perturbative BRST symmetry enables us to prove that the correlation functions of composite operators belonging to the cohomology of the BRST operator turn out to be independent from the  gauge parameter.

By taking into account further non-perturbative effects, a refinement of the Gribov-Zwanziger action was proposed \cite{Dudal:2008sp}. It arises from the inclusion of dynamical effects such as the formation of dimension-two condensates in the original Gribov-Zwanziger action. The ensuing  Refined Gribov-Zwanziger gluon and ghost propagators are in qualitative agreement with the most recent lattice data \cite{Cucchieri:2007rg,Cucchieri:2008fc,Cucchieri:2011ig,Maas:2008ri}. Also, developments as the inclusion of matter \cite{Capri:2014bsa}, glueball spectrum \cite{Dudal:2010cd,Dudal:2013wja}, thermodynamics \cite{Canfora:2015yia,Canfora:2013kma,Canfora:2013zna,Lichtenegger:2008mh,Fukushima:2012qa,Fukushima:2013xsa,Guimaraes:2015vra} and generalization to supersymmetric theories \cite{Capri:2014xea,Capri:2014tta} were made.

The gluon propagator obtained from the Refined Gribov-Zwanziger action in the Landau gauge has the so-called massive (decoupling) behavior in $D=3,4$ and a scaling behavior in $D=2$, \cite{Dudal:2008sp,Dudal:2008rm,Dudal:2008xd}. A massive behavior is characterized by a suppression in the infrared regime and a finite non-vanishing value at zero momentum, while a scaling behavior accounts also for a suppression, but a vanishing value at zero momentum. These results are in qualitative agreement with lattice data. Also, the Refined Gribov-Zwanziger framework was studied in the Coulomb and maximal Abelian gauges sharing these properties of the Landau gauge. In the Coulomb gauge, a scaling behavior for the spatial propagator at equal time was observed in $D=3$, while a decoupling behavior for the same propagator was found in $D=4$, see \cite{Guimaraes:2015bra}. In the maximal Abelian gauge, the diagonal gluon propagator turns out to  be of the scaling type in $D=2$ and decoupling in $D=3,4$, see \cite{Capri:2015pfa} and \cite{Amemiya:1998jz,Bornyakov:2003ee,Mendes:2006kc,Gongyo:2013sha,Gongyo:2014lxa,Schrock:2015pna} for various lattice studies over the years.

In recent years, steady progress has been made on the non-perturbative study of the linear covariant gauges within different approaches, see for instance \cite{Cucchieri:2009kk,Bicudo:2015rma,Siringo:2014lva,Siringo:2015gia,Huber:2015ria,Aguilar:2015nqa,Aguilar:2007nf}.
The linear covariant gauges arise thus as a fertile terrain for further development of non-perturbative analytical frameworks through the active comparison with lattice simulations and functional methods. The goal of the present paper is to discuss the predictions for the gluon propagator in the Gribov-Zwanziger framework for linear covariant gauges in different Euclidean spacetime dimensions ($D=2,3$ and $4$). With that aim, the dynamical generation of dimension-two condensates has to be addressed in each case, in order to obtain the correct prediction for the infrared (IR) behavior of the gluon correlator.

In \cite{Sobreiro:2005vn,Capri:2015pja}, a construction of the Gribov-Zwanziger action for linear covariant gauges was proposed.
A further generalization which exhibits manifest non-perturbative BRST invariance was presented in \cite{Capri:2015ixa}. The non-perturbative invariance controls the gauge-parameter dependence in a very explicit way, which makes the construction consistent with gauge invariance. Also, the modified BRST symmetry protects the longitudinal component of the gluon propagator\footnote{This is trivial if we assume that the perturbative BRST symmetry remains at the non-perturbative level. However, if the perturbative BRST symmetry is broken - which is the case in the Gribov-Zwanziger approach - we cannot ensure it from the beginning.}  from  non-perturbative modifications, while the transverse part receives modifications similar to the Landau ones. In this way, we are able to compute the gluon propagator and conclude that the Landau behavior, \textit{i.e.}, massive/decoupling in $D=3,4$ and scaling in $D=2$ is preserved, when the inclusion of condensates is taken into account.

The paper is organized as follows: In Sect.~\ref{GZnpBRST}, we provide a review of the construction of the non-perturbative BRST symmetry for the Gribov-Zwanziger action in the Landau gauge. In Sect.~\ref{LCGext}, we extend our results to  continuum linear covariant gauges, through the introduction of an action which restricts the path integral domain to a Gribov region which is free of a large set of gauge copies. This action is manifestly invariant under a non-perturbative BRST transformation and shares many features with the Gribov-Zwanziger action in the Landau gauge. In Sect.~\ref{GZnplatt}, we discuss a possible lattice construction of the non-perturbative continuum formulation used in this paper and its predecessor \cite{Capri:2015ixa}.  We then argue that the Gribov-Zwanziger vacuum in the linear covariant gauge develops further non-perturbative effects via the generation of mass dimension 2 condensates. Therefore, we extend the results of Sect.~\ref{LCGext} by the introduction of such condensates in the action, akin to the Refined Gribov-Zwanziger action. This introduction is consistent in $D=3,4$, while in $D=2$, a non-integrable IR singularity prevents their introduction. We also discuss how a similar IR obstruction would lead to an inconsistency with the restriction to the Gribov region in the presence of the condensate in $D=2$.  These results are discussed in Sect.~\ref{RGZ}. In Sect.~\ref{Gprop}, we show the gluon propagator in $D=3,4$ and in $D=2$. For the former cases, we observe a decoupling behavior while for the latter, a scaling one. Finally, we point out our conclusions.

\section{The Gribov-Zwanziger framework in the Landau gauge and the non-perturbative BRST symmetry} \label{GZnpBRST}
The Gribov-Zwanziger action in the Landau gauge, for $SU(N)$ gauge group and in $D$-dimensional Euclidean space, implements the restriction of the path integral domain to the Gribov region $\Omega$, defined as
\begin{equation}
\Omega=\left\{A^{a}_{\mu}\,\,|\,\,\partial_{\mu}A^{a}_{\mu}=0\,,\,\,\,\EuScript{M}^{ab}>0\right\}\,,
\label{landau1}
\end{equation}
where $\EuScript{M}^{ab}$ is the Faddeev-Popov operator in the Landau gauge,
\begin{equation}
\EuScript{M}^{ab}(A)\equiv -\partial_{\mu}D^{ab}_{\mu}=-\delta^{ab}\partial^2+gf^{abc}A^{c}_{\mu}\partial_{\mu}\,,
\label{landau2}
\end{equation}
with the Landau gauge condition $\partial_{\mu}A^{a}_{\mu}=0$ imposed. The operator $\EuScript{M}^{ab}(A)$ is hermitian, which is a very important property for constructing $\Omega$, a region where $\EuScript{M}^{ab}(A)$ is strictly positive and, therefore, does not develop any zero modes. This region is free from infinitesimal Gribov copies, which are related to zero modes $\omega$ of the Faddeev-Popov operator. Indeed, assuming that
\begin{equation}
\EuScript{M}^{ab}\omega^b=0\,,
\end{equation}
then we immediately see that also
\begin{equation}
\p_\mu(A_\mu^a+D_\mu^{ab}\omega^b)=0\,,
\end{equation}
so that a gauge copy is associated  to each zero mode.

Remarkably, the region $\Omega$ enjoys a large set of properties which makes the restriction to $\Omega$ a consistent procedure: it is bounded in all directions, convex and all gauge orbits cross it at least once, \cite{Dell'Antonio:1991xt}. We should emphasize that although free from a large set of copies, $\Omega$ still contains Gribov copies (related through finite gauge transformations), see \cite{vanBaal:1991zw}. The restriction implemented by Gribov and generalized by Zwanziger is obtained via
\begin{equation}
\EuScript{Z}=\int_{\Omega}\left[\EuScript{D}A\right]\mathrm{det}\left(\EuScript{M}\right)\delta\left(\partial A\right)\mathrm{e}^{-S_{\mathrm{YM}}}=\int\left[\EuScript{D}A\right]\mathrm{det}\left(\EuScript{M}\right)\delta\left(\partial A\right)\mathrm{e}^{-\left(S_{\mathrm{YM}}+\gamma^4H(A)-DV\gamma^4(N^2-1)\right)}\,,
\label{landau3}
 \end{equation}
where
\begin{equation}
H(A)=g^2\int \d^D x\d^Dy~f^{abc}A^{b}_{\mu}(x)\left[\EuScript{M}^{-1}(x,y)\right]^{ad}f^{dec}A^{e}_{\mu}(y)
\label{landau4}
\end{equation}
is the so-called horizon function and $V$ stands for the spacetime volume. It involves a mass parameter $\gamma$, known as the Gribov parameter, which is not free but fixed by the gap equation
\begin{equation}
\langle H \rangle = DV(N^2-1)\,,
\label{landau5}
\end{equation}
where vacuum expectation values $\langle \ldots \rangle$ are taken with the measure defined in eq.(\ref{landau3}). It will set $\gamma\propto \Lambda_{\textrm{QCD}}$ via dimensional transmutation and as such the Gribov parameter becomes a non-perturbative mass scale.

The Gribov-Zwanziger action is defined as
\begin{equation}
\tilde{S}_{\mathrm{GZ}}=S_{\mathrm{FP}}+\gamma^4H(A)\,,
\label{landau6}
\end{equation}
where $S_{\mathrm{FP}}$ corresponds to the usual Faddeev-Popov action in the Landau gauge,
\begin{equation}
S_{\mathrm{FP}}=S_{\mathrm{YM}}+\int \d^D x~\left(b^a\partial_{\mu}A^{a}_{\mu}+\bar{c}^{a}\partial_{\mu}D^{ab}_{\mu}c^b\right)\,,
\label{landau7}
\end{equation}
with $(\bar{c},c)$ standing for the Faddeev-Popov ghosts and $b$ is the auxiliary Nakanishi-Lautrup field. However, in this form, the Gribov-Zwanziger action is non-local, due to the presence of the horizon function. Nevertheless, it turns out that it is possible to recast this action in a local form by the introduction of a set of auxiliary fields, namely a pair of commuting fields, $(\bar{\varphi},\varphi)$, and a pair of anticommuting ones, $(\bar{\omega},\omega)$. Therefore, the Gribov-Zwanziger action is written as
\begin{equation}
S_{\mathrm{GZ}}=S_{\mathrm{FP}}+\int \d^Dx~\left(\bar{\varphi}^{ac}_{\mu}\EuScript{M}^{ab}(A)\varphi^{bc}_{\mu}-\bar{\omega}^{ac}_{\mu}\EuScript{M}^{ab}(A)\omega^{bc}_{\mu}+g\gamma^2f^{abc}A^{a}_{\mu}(\varphi+\bar{\varphi})^{bc}_{\mu}\right)\,.
\label{landau8}
\end{equation}
In this local form, the gap equation (\ref{landau5}) is given by
\begin{equation}
\frac{\partial\mathcal{E}_v}{\partial\gamma^2}=0\,,\,\,\,\,\,\,\, \mathrm{with}\,\,\,\,\,\,\,\mathrm{e}^{-V\mathcal{E}_v}=\int \left[\EuScript{D}\Phi\right]\mathrm{e}^{-\left(S_{\mathrm{GZ}}-DV\gamma^4(N^2-1)\right)}\,,
\label{landau9}
\end{equation}
where $\Phi$ denotes the entire set of fields and $\mathcal{E}_v$, the vacuum energy.

The Gribov-Zwanziger action (\ref{landau8}) is local and renormalizable to all orders in perturbation theory \cite{Zwanziger:1992qr,Maggiore:1993wq,Dudal:2008sp,Dudal:2010fq,Capri:2015mna}. However, it breaks BRST softly, a point extensively discussed in the literature, see \cite{Capri:2014bsa,Dudal:2009xh,Sorella:2009vt,Baulieu:2008fy,Capri:2010hb,Dudal:2012sb,Dudal:2014rxa,Capri:2012wx,Pereira:2013aza,Pereira:2014apa,Tissier:2010ts,Serreau:2012cg,Serreau:2015yna,Lavrov:2013boa,Moshin:2015gsa,Schaden:2014bea,Cucchieri:2014via}. This breaking can be naively expected to be a consequence of the restriction of the domain of the path integral, as we are enforcing a non-trivial boundary in field space in which the BRST transformation connects different configurations \cite{Dudal:2014rxa}. Therefore, it sounds reasonable to wonder if a suitable modification of the usual BRST transformations to a new one that feels the restriction of the path integral to $\Omega$ and that corresponds to a symmetry of the Gribov-Zwanziger action is possible. Such construction was addressed in a recent paper \cite{Capri:2015ixa} and the modified BRST transformation, called non-perturbative BRST, leads to a symmetry of the Gribov-Zwanziger action.

The non-perturbative BRST transformation is constructed through the introduction of the non-local field $A^h_{\mu}$,
\begin{equation}
A_{\mu }^{h} =\left( \delta _{\mu \nu }-\frac{\partial _{\mu }\partial
_{\nu }}{\partial ^{2}}\right) \left( A_{\nu }-ig\left[ \frac{1}{\partial
^{2}}\partial A,A_{\nu }\right] +\frac{ig}{2}\left[ \frac{1}{\partial ^{2}}%
\partial A,\partial _{\nu }\frac{1}{\partial ^{2}}\partial A\right] \right)
+O(A^{3})\,,
\label{landau10}
\end{equation}
where we are employing the matrix notation $A_{\mu}=A^{a}_{\mu}T^{a}$, with $T^a$ denoting the hermitian $SU(N)$ generators satisfying the algebra $[T^a,T^b]=if^{abc}T^c$. The field $A^h$ is transverse and gauge-invariant order by order, a very important property underpinning the construction. Let us refer to \cite{Lavelle:1995ty,Capri:2015ixa} for the derivation of expression \eqref{landau10}. Notice that this is a series expansion, so questions might be raised about its convergence for larger coupling/gauge fields. We will come back to this issue in Sect.~\ref{GZnplatt}.

We notice that, in eq.(\ref{landau10}), all terms but the ones linear in $A_{\mu}$ contain at least one factor of $\partial A$ which allows us to write the horizon function in the form
\begin{equation}
H(A)=H(A^h)-R(A)(\partial A)\,,
\label{landau11}
\end{equation}
with $R(A)(\partial A)=\int \d^Dx\d^Dy~R^a(x,y)(\partial A^a)_y$, a non-local expression which contains contributions proportional to $(\partial A)$ in (\ref{landau10}). It implies we can rewrite the Gribov-Zwanziger action as
\begin{equation}
\tilde{S}_{\mathrm{GZ}}=S_{\mathrm{YM}}+\int \d^Dx\left(b^{h,a}\partial_{\mu}A^{a}_{\mu}+\bar{c}^{a}\partial_{\mu}D^{ab}_{\mu}c^b\right)+\gamma^4H(A^h)\,,
\label{landau12}
\end{equation}
where $b^{h,a}$ is a redefinition of the $b^a$ field with trivial Jacobian, given by
\begin{equation}
b^{h,a}=b^a-\gamma^4R(A)\,.
\label{landau13}
\end{equation}
Introducing the auxiliary Zwanziger's fields $(\bar{\varphi},\varphi,\bar{\omega},\omega)$, the reformulated Gribov-Zwanziger action reads
\begin{eqnarray}
S_{\mathrm{GZ}}&=& S_{\mathrm{YM}}+\int \d^Dx\left(b^{h,a}\partial_{\mu}A^{a}_{\mu}+\bar{c}^{a}\partial_{\mu}D^{ab}_{\mu}c^{b}\right)\nonumber\\
&+&\int \d^Dx\left(\bar{\varphi}^{ac}_{\mu}\left[\EuScript{M}(A^h)\right]^{ab}\varphi^{bc}_{\mu}-\bar{\omega}^{ac}_{\mu}\left[\EuScript{M}(A^h)\right]^{ab}\omega^{bc}_{\mu}+g\gamma^2f^{abc}A^{h,a}_{\mu}(\varphi+\bar{\varphi})^{bc}_{\mu}\right)\,,
\label{landau14}
\end{eqnarray}
which enjoys the new non-perturbative BRST symmetry,
\begin{align}
s_{\gamma^2}A^{a}_{\mu}&=-D^{ab}_{\mu}c^b\,,     &&s_{\gamma^2}c^a=\frac{g}{2}f^{abc}c^bc^c\,, \nonumber\\
s_{\gamma^2}\bar{c}^a&=b^{h,a}\,,     &&s_{\gamma^2}b^{h,a}=0\,, \nonumber\\
s_{\gamma^2}\varphi^{ab}_{\mu}&=\omega^{ab}_{\mu}\,,   &&s_{\gamma^2}\omega^{ab}_{\mu}=0\,, \nonumber\\
s_{\gamma^2}\bar{\omega}^{ab}_{\mu}&=\bar{\varphi}^{ab}_{\mu}+\gamma^2gf^{cdb}A^{h,c}_{\mu}\left[\EuScript{M}^{-1}(A^h)\right]^{da}\,,         &&s_{\gamma^2}\bar{\varphi}^{ab}_{\mu}=0\,, \nonumber\\
s_{\gamma^2}A^{h,a}_{\mu}&=0\,.
\label{landau15}
\end{align}
The operator $s_{\gamma^2}$ is nilpotent, $s_{\gamma^2}^2=0$, and corresponds to an exact symmetry of (\ref{landau14}). We must comment that (\ref{landau14}) is a non-local expression as well as the transformations (\ref{landau15}). It would be highly desirable to cast this framework in a local form, a subject already under investigation \cite{Capri}. However, we can already extract important information from these non-local expressions for our present purposes.

Before continuing, it is perhaps important to remind that at the level of the gauge field itself, $s_{\gamma^2}$ still implements nothing more than an infinitesimal local gauge transformation. This will be important to make the connection between classically gauge-invariant operators and their full BRST-invariant quantum counterparts. The non-perturbative nature of the new BRST operator resides in its explicit dependence on $\gamma\propto \Lambda_{\textrm{QCD}}$. Also, the non-perturbative nature of $\gamma$ ensures that in the UV region the $\gamma$-dependent BRST transformations simply reduce to the standard perturbative one.

\section{Extension to the linear covariant gauges} \label{LCGext}
A big challenge for the Gribov-Zwanziger approach has always been a proper generalization of the framework to different gauges. Besides the Landau choice, important developments were achieved in the maximal Abelian gauge \cite{Gongyo:2013rua,Capri:2015pfa} and in the Coulomb gauge \cite{Gribov:1977wm,Guimaraes:2015bra,Burgio:2008jr,Burgio:2009xp,Zwanziger:2004np}. The main reason is that in the Landau gauge, the Faddeev-Popov operator, whose zero modes correspond to infinitesimal Gribov copies, is hermitian. This feature allows us to define a region where such operator is strictly positive due to its real spectrum. However, if we relax this condition, we immediately loose this geometric interpretation and the construction of a region to restrict the path integral domain becomes unclear.

The extension of the Gribov-Zwanziger framework to linear covariant gauges faces precisely this problem. In this class of gauges, the Faddeev-Popov operator is not hermitian due to the fact the gauge field is not transverse on-shell. Some attempts to the extension of the Gribov-Zwanziger action to linear covariant gauges were done in \cite{Sobreiro:2005vn,Capri:2015pja}. In this section, we extend the formalism of the last section to linear covariant gauges, which provides a Gribov-Zwanziger action which enjoys the non-perturbative symmetry and implements the restriction of the domain of integration to a region which is free of a large set of Gribov copies, see \cite{Capri:2015ixa}. The extension of the Gribov-Zwanziger action to linear covariant gauges is
\begin{eqnarray}
S^{\mathrm{LCG}}_{\mathrm{GZ}}&=& S_{\mathrm{YM}}+s_{\gamma^2}\int \d^Dx~\bar{c}^{a}\left(\partial_{\mu}A^{a}_{\mu}-\frac{\alpha}{2}b^{h,a}\right)\nonumber\\
&+&\int \d^Dx\left(\bar{\varphi}^{ac}_{\mu}\left[\EuScript{M}(A^h)\right]^{ab}\varphi^{bc}_{\mu}-\bar{\omega}^{ac}_{\mu}\left[\EuScript{M}(A^h)\right]^{ab}\omega^{bc}_{\mu}+g\gamma^2f^{abc}A^{h,a}_{\mu}(\varphi+\bar{\varphi})^{bc}_{\mu}\right)\nonumber\\
&=&  S_{\mathrm{YM}} + \int \d^Dx\left(b^{h,a}\left(\partial_{\mu}A^{a}_{\mu}-\frac{\alpha}{2}b^{h,a}\right)+\bar{c}^{a}\partial_{\mu}D^{ab}_{\mu}c^{b}\right)\nonumber\\
&+&\int d^Dx\left(\bar{\varphi}^{ac}_{\mu}\left[\EuScript{M}(A^h)\right]^{ab}\varphi^{bc}_{\mu}-\bar{\omega}^{ac}_{\mu}\left[\EuScript{M}(A^h)\right]^{ab}\omega^{bc}_{\mu}+g\gamma^2f^{abc}A^{h,a}_{\mu}(\varphi+\bar{\varphi})^{bc}_{\mu}\right)\,,
\label{lcg1}
\end{eqnarray}
where the gauge condition for linear covariant gauges is defined as
\begin{equation}
\partial_{\mu}A^{a}_{\mu}-\alpha b^{h,a}=0\,,
\label{lcg2}
\end{equation}
with $\alpha$ being an arbitrary positive parameter. Due to the nilpotency of $s_{\gamma^2}$, the action (\ref{lcg1}) is manifestly invariant under non-perturbative BRST transformations. As discussed in \cite{Capri:2015ixa}, this action restricts the domain of integration of the path integral to a region which is free of a large set of Gribov copies.

More precisely, it restricts the path integral to gauge fields $A_\mu$ that fulfill the gauge condition, next to ensuring positivity of\footnote{Notice that the argument of the Faddeev-Popov operator here is the gauge-invariant transverse field $A_\mu^h$. As such, this operator is again hermitian, but it does not equal the Faddeev-Popov operator of the linear covariant gauges.} $\EuScript{M}^{ab}(A^h)$. As explained in detail in \cite{Capri:2015ixa,Capri:2015pja}, this ensures the absence of zero modes $\omega$ of the Faddeev-Popov operator $\EuScript{M}^{ab}(A)$ itself, modulo the assumption that $\omega$  is Taylor-expandable around $\alpha=0$. A fortiori, the gauge copies associated with these zero modes are then also eliminated.

We must mention that in previous works \cite{Sobreiro:2005vn,Capri:2015pja} an action akin to the Gribov-Zwanziger action was obtained through the imposition of the restriction of the transverse gauge field to the Gribov region $\Omega$. The resulting action is equivalent to (\ref{lcg1}) in the lowest-order approximation
\begin{equation}
A^h_{\mu}\approx A_{\mu}-\frac{\partial_{\mu}}{\partial^2}(\partial A)\equiv A^{T}_{\mu}\,.
\label{lcg3}
\end{equation}
The invariance under non-perturbative BRST transformations, however, brings a much clearer scenario. This symmetry provides a powerful tool to control the dependence from $\alpha$ of correlation functions of gauge-invariant composite operators. In particular, the gap equation  which fixes the Gribov parameter $\gamma$ can be written as
\begin{equation}
\langle H(A^h)\rangle = DV(N^2-1)\,,
\label{lcg4}
\end{equation}
a manifestly gauge-invariant expression which implements the gauge independence of $\gamma$ and, thus, attributes to it a genuine physical meaning. As a consequence, it can enter expectation values of gauge-invariant operators $\mathcal{O}_{\textrm{g inv}}$ which will obey $s_{\gamma^2}\mathcal{O}_{\textrm{g inv}}=0$. Moreover, given that the gauge parameter $\alpha$ is coupled to a $s_{\gamma^2}$-exact form in $S^\textrm{LCG}_{\textrm{GZ}}$, it holds that $\frac{\d}{\d \alpha}\braket{\mathcal{O}_{\textrm{g inv}}(x)  \mathcal{O}_{\textrm{g inv}}(y)}=0$. A general proof of this outcome is a subject of another work of the present authors.

\section{A minimizing functional for the non-perturbative linear covariant gauge}\label{GZnplatt}
In the current paper, based on \cite{Capri:2015ixa}, a pivotal role in the whole construction is played by the transverse gauge-invariant field $A_\mu^h$, introduced in eq.~\eqref{landau10}. As noticed before, the  explicit expression for $A_\mu^h$ is a formal power series expansion in the original field $A_\mu$, so it becomes difficult to access $A^h$ if such power series would be needed to (arbitrarily) large orders. We recall here that $A_\mu^h$ is constructed via the minimization of the functional $\int \d^Dx A_\mu^a A_\mu^a$ along the gauge orbit, see e.g.~\cite{Capri:2015ixa}. A local minimum is achieved when $A_\mu^a$ would be transverse and its associated Faddeev-Popov operator positive. This is also nothing else than the way how the Landau gauge is implemented numerically, i.e.~via a minimization procedure of a suitable functional for which powerful algorithm exists, see e.g.~\cite{Maas:2015nva,Cucchieri:2013nja} for recent interesting works.

For future lattice numerical studies, it would thus also be interesting to have at our disposal a minimizing functional that would implement the linear covariant gauge as analytically investigated here. Minimizing functionals for a lattice version of the linear covariant gauge were already introduced in \cite{Cucchieri:2009kk} and used in \cite{Bicudo:2015rma}. The importance of minimizing functionals was also recognized in \cite{Serreau:2015yna}. The functionals in \cite{Cucchieri:2009kk,Bicudo:2015rma} do lead to the gauge condition \eqref{lcg2} via the vanishing of the first derivative, though the second derivative is not really used to our understanding.  For our definition of the linear covariant gauge, we also need to ensure $\EuScript{M}^{ab}(A^h)>0$.

Therefore, consider the (positive) functional
\begin{equation}\label{fg1}
    \mathcal{R}(A,B,U,V)\equiv  \text{Tr} \int \d^Dx \left(A_\mu^UA_\mu^U+\frac{2}{g}\text{Re}(i U\Lambda)\right)+  \text{Tr} \int \d^Dx \left(B_\mu^V B_\mu^V\right)+ \text{Tr} \int \d^Dx \left( B_\mu^V- P_{\mu\nu}A_\nu^U\right)^2\,,
\end{equation}
with $P_{\mu\nu}=\delta_{\mu\nu}-\frac{\p_\mu \p_\nu}{\p^2}$ the usual transversal projector.

We define our gauge by looking at the (local) minima of $\mathcal{R}(A,B,U,V)$, for a fixed function $\Lambda(x)= \Lambda^a(x)t^a$ in function of variable $U,V$. We work with conventions $[t^a,t^b]=if^{abc} t^c $, $ \text{Tr}(t^a t^b)=\frac{1}{2}$, $A_\mu^U= U^\dagger A_\mu U + \frac{i}{g}U^\dagger \p_\mu U$ and similarly for $B_\mu^V$ and $V$.

We will compute the variation of $\mathcal{R}(A,B,U,V)$ up to second-order in $\omega,\theta$, with $U=e^{ig\omega^a t^a}$, $V=e^{ig\theta^a t^a}$ generic gauge transformations.

Clearly, the third term of $\mathcal{R}$ (cf.~\eqref{fg1}) will be minimal (zero) if
\begin{equation}\label{fg2}
  B_\mu= P_{\mu\nu} A_\nu
\end{equation}
thus $B_\mu$ is the transverse part of $A_\mu$.

Let us now consider the variation of the first and second terms of $\mathcal{R}$,
\begin{eqnarray}\label{fg2b}
&&\int \d^Dx \left(A_\mu^UA_\mu^U+\frac{2}{g}\text{Re}(iU\Lambda)\right)+\int \d^Dx \left(B_\mu^VB_\mu^V\right) = \int \d^Dx \left(A_\mu^a A_\mu^a+B_\mu^a B_\mu^a\right)\nonumber\\
&&+\int \d^Dx \left(\omega^a \p_\mu A_\mu^a-\omega^a \Lambda^a+\theta^a \p_\mu B_\mu^a\right)-\frac{1}{2}\int \d^Dx\left[\omega^a \left(\p_\mu D_\mu^{A,ab}+D_\mu^{A,ab} \p_\mu\right)\omega^b\right]\nonumber\\&&-\frac{1}{2}\int \d^Dx\left[\theta^a \left(\p_\mu D_\mu^{B,ab}+D_\mu^{B,ab} \p_\mu\right)\theta^b\right]+\ldots
\end{eqnarray}
The notation $D_\mu^{X,ab}$ refers to the covariant derivative w.r.t.~$X=A,B$.

So, the first-order variations vanish when
\begin{equation}\label{fg3}
    \p_\mu A_\mu^a=\Lambda^a\,,\qquad \p_\mu B_\mu^a=0
\end{equation}
The second condition is however obsolete since we already have \eqref{fg2}. In any case, our gauge field $A_\mu$ is confined to the linear covariant gauge.  This gauge clearly leaves the transverse part of $A_\mu$, viz.~$B_\mu$, undetermined.

Likewise, the second-order variations will be positive if\footnote{We already used that $B_\mu$ is transverse here.}
\begin{equation}\label{fg3b}
    -\frac{1}{2}(\p_\mu D_\mu^{A,ab}+D_\mu^{A,ab} \p_\mu)>0,\qquad  -\p_\mu D_\mu^{B,ab}>0
\end{equation}
By construction, $B_\mu$ is nothing else than $A_\mu^h$: it corresponds to a (local) minimum of $\int B^2$. The trick is that we \emph{first} identified $B_\mu$ with the transverse part of the original gauge field $A_\mu$, which is acceptable since $A_\mu^h$ is transverse by construction. So, $B_\mu \equiv A_\mu^h$ and we have assured that $\EuScript{M}(A^h)>0$.

There is one extra condition, namely the positivity of the hermitian operator $\Delta\equiv-(\p_\mu D_\mu^{A,ab}+D_\mu^{A,ab} \p_\mu)$.
We notice that, employing generic real test functions $\rho$,
\begin{eqnarray}\label{pos}
\Delta> 0 &\Leftrightarrow& -\frac{1}{2}\int \d^Dx \rho^a(\p_\mu D_\mu^{ab}+D_\mu^{ab} \p_\mu)\rho^b>0\nonumber\\
&\Leftrightarrow&  -\int \d^Dx \rho^a(\p_\mu D_\mu^{ab})\rho^b - \underbrace{\frac{1}{2}\int \d^Dx \rho^a gf^{abc}\rho^b \p_\mu A_\mu^c}_{=0\text{ due to antisymmetry of $f^{abc}$}}>0 ~\Leftrightarrow~ ``-\p D > 0"
\end{eqnarray}
The last identification is however a bit formal, since to call an operator positive one usually requires the operator to have only positive eigenvalues: a priori, only for hermitian operators the eigenvalues are real, which is of course a necessity to check whether they are positive. The operator $-\p D$ is however not hermitian, contrasting with $-\p D-D\p$.

However, the reasoning \eqref{pos} is useful anyhow. We recall that $\p A=\p A'=\Lambda$ could occur for infinitesimally connected equivalent gauge fields $A$ and $A'$ if and only if $\p D \omega=0$, i.e.~if $\p D$ has zero modes. Silently, we can restrict to real-valued $\omega$, since the operators are supposed to act on real spaces, the reason being that after all, the $\omega$ are referring to the transformation variables (``angles'') of $SU(N)$ transformations. The standard Faddeev-Popov operator itself is also obtained via the action of local $SU(N)$ gauge rotations, that is, with real transformation variables.

To proceed, assuming that $\omega$ is such a real zero mode, $-\p D \omega=0$, a fortiori one has $-\int\omega \p D \omega=0$, which would be inconsistent with $(-\omega \p D \omega~=)~\omega\Delta\omega>0$. So, requiring $\Delta>0$ (in the real space it acts on) is formally equivalent to ``$-\p D>0$'' and also takes care of the (partial) resolution of the Gribov problem, as infinitesimal gauge copies are excluded. But in \cite{Capri:2015ixa}, we actually showed that for smooth copies, imposing $\EuScript{M}(A^h)>0$ already kills off the infinitesimal copies, so it is not even necessary to further impose that $\Delta>0$. From this perspective, it would seem that a numerical minimization of \eqref{fg1} can be used to implement our desired non-perturbative linear covariant gauge.

\section{Introducing further dynamical effects: the non-perturbative BRST-invariant Refined Gribov-Zwanziger action} \label{RGZ}

In the Landau gauge, it was shown that the Gribov-Zwanziger action suffers from instabilities that give rise to the dynamical generation of condensates. Such effects can be taken into account by the construction of the so-called Refined Gribov-Zwanziger action, \cite{Dudal:2007cw,Dudal:2008sp,Gracey:2010cg}. The same issue was analyzed in the maximal Abelian gauge, \cite{Capri:2015pfa} and in the Coulomb gauge \cite{Guimaraes:2015bra}. In \cite{Capri:2015pja}, the existence of non-vanishing condensates at one-loop in linear covariant gauges was pointed out, in the framework of (\ref{lcg3}).

Clearly, at one-loop order, the computation using the action (\ref{lcg1}) or the one using the expression (\ref{lcg3}) leads to the same results for the condensates. However, we should keep in mind these formulations are conceptually very different. In the present case, we are concerned with the condensates

\begin{equation}
\langle A^{h,a}_{\mu}A^{h,a}_{\mu}\rangle\,\,\,\,  \mathrm{and}\,\,\,\, \langle \bar{\varphi}^{ab}_{\mu}\varphi^{ab}_{\mu}-\bar{\omega}^{ab}_{\mu}\omega^{ab}_{\mu}\rangle\,,
\label{rgz1}
\end{equation}
while in \cite{Capri:2015pja}, the dimension-two gluon condensate was $\langle A^{Ta}_{\mu}A^{Ta}_{\mu}\rangle$. Explicitly, the value of the condensates (\ref{rgz1}) can be computed by coupling such composite operators with constant sources $m$ and $J$, namely

\begin{equation}
\mathrm{e}^{-V\mathcal{E}(m,J)}=\int\left[\EuScript{D}\Phi\right]\mathrm{e}^{-(S^{\mathrm{LCG}}_{\mathrm{GZ}}+m\int \d^Dx~A^{h,a}_{\mu}A^{h,a}_{\mu}-J\int d^Dx(\bar{\varphi}^{ab}_{\mu}\varphi^{ab}_{\mu}-\bar{\omega}^{ab}_{\mu}\omega^{ab}_{\mu}))}
\label{rgz2}
\end{equation}
and
\begin{eqnarray}
\langle \bar{\varphi}^{ab}_{\mu}\varphi^{ab}_{\mu}-\bar{\omega}^{ab}_{\mu}\omega^{ab}_{\mu} \rangle &=& -\frac{\partial \mathcal{E}(m,J)}{\partial J}\Big|_{m=J=0}\nonumber\\
\langle A^{h,a}_{\mu}A^{h,a}_{\mu}\rangle &=& \frac{\partial\mathcal{E}(m,J)}{\partial m}\Big|_{m=J=0}\,.
\label{rgz3}
\end{eqnarray}
At one-loop order, employing dimensional regularization,
\begin{equation}
{\cal E}(m,J)=\frac{(D-1)(N^2-1)}{2}\int \frac{d^Dk}{(2\pi)^D}~\mathrm{ln}\left(k^2+\frac{2\gamma^4g^2N}{k^2+J}+2m\right)-D\gamma^4(N^2-1)\,,
\label{rgz4}
\end{equation}
which results in
\begin{equation}
\langle \bar{\varphi}^{ac}_{\mu}\varphi^{ac}_{\mu}-\bar{\omega}^{ac}_{\mu}\omega^{ac}_{\mu}\rangle = g^2\gamma^4N(N^2-1)(D-1)\int \frac{d^Dk}{(2\pi)^D}\frac{1}{k^2}\frac{1}{(k^4+2g^2\gamma^4N)}
\label{rgz5}
\end{equation}
and
\begin{equation}
\langle A^{h,a}_{\mu}A^{h,a}_{\mu}\rangle = -2g^2\gamma^4N(N^2-1)(D-1)\int\frac{\d^Dk}{(2\pi)^D}\frac{1}{k^2}\frac{1}{(k^4+2g^2\gamma^4N)}\,.
\label{rgz6}
\end{equation}
From (\ref{rgz5}) and (\ref{rgz6}), we see the integrals are perfectly convergent in the UV and depend explicitly on $\gamma$. For $D=3,4$, these integrals are defined in the IR and correspond to well-defined quantities. Nevertheless, in $D=2$, due to the $1/k^2$ factor in the integrals, we have a non-integrable singularity which makes the condensates ill-defined. This IR pathology in $D=2$ is a typical behavior of two-dimensional theories, see \cite{Dudal:2008xd} and references therein. In this way, these results suggest such condensates should be taken into account in $D=3,4$, giving rise to a refinement of the Gribov-Zwanziger action. In $D=2$, as happens in other gauges, these condensates cannot be safely introduced as they give rise to non-integrable IR singularities. As a consequence, in $D=2$ the Gribov-Zwanziger theory does not need to be refined.  Therefore, for $D=3,4$, the Refined Gribov-Zwanziger action in linear covariant gauges is written as
\begin{eqnarray}
S^{\mathrm{LCG}}_{\mathrm{RGZ}}&=& S_{\mathrm{YM}} + \int \d^Dx\left(b^{h,a}\left(\partial_{\mu}A^{a}_{\mu}-\frac{\alpha}{2}b^{h,a}\right)+\bar{c}^{a}\partial_{\mu}D^{ab}_{\mu}c^{b}\right)\nonumber\\
&+&\int \d^Dx\left(\bar{\varphi}^{ac}_{\mu}\left[\EuScript{M}(A^h)\right]^{ab}\varphi^{bc}_{\mu}-\bar{\omega}^{ac}_{\mu}\left[\EuScript{M}(A^h)\right]^{ab}\omega^{bc}_{\mu}+g\gamma^2f^{abc}A^{h,a}_{\mu}(\varphi+\bar{\varphi})^{bc}_{\mu}\right)\nonumber\\
&+&\frac{m^2}{2}\int \d^Dx~A^{h,a}_{\mu}A^{h,a}_{\mu}-M^2\int \d^Dx\left(\bar{\varphi}^{ac}_{\mu}\varphi^{ac}_{\mu}-\bar{\omega}^{ac}_{\mu}\omega^{ac}_{\mu}\right)\,,
\label{rgz7}
\end{eqnarray}
while for $D=2$, the action is simply the Gribov-Zwanziger action, given by eq.(\ref{lcg1}). Notice that $M^2\geq0$, otherwise the theory would be plagued by a tachyon in the $(\omega,\bar\omega)$-sector.

The action (\ref{rgz7}) enjoys a non-perturbative nilpotent BRST symmetry, which is precisely the same as (\ref{landau15}) with the only modification of
\begin{equation}
s_{\gamma^2}\bar{\omega}^{ab}_{\mu}=\bar{\varphi}^{ab}_{\mu}+g\gamma^2f^{cdb}A^{h,c}_{\mu}\left(\left[\EuScript{M}(A^h)- \mathds{1} M^2 \right]^{-1}\right)^{da}\,,
\label{rgz8}
\end{equation}
where $\mathds{1}$ stands for the identity operator. Therefore, the Refined Gribov-Zwanziger action in linear covariant gauges takes into account the presence of dimension-two condensates and is invariant under (\ref{landau15}) and (\ref{rgz8}), a non-perturbative nilpotent BRST symmetry.

There is an additional problem, of a more fundamental nature, that prohibits the dynamical occurrence of refinement in $D=2$. We recall here from the earlier Sect.~\ref{GZnpBRST} that the starting point was to avoid a large class of infinitesimal gauge copies in the linear covariant gauge. This was achieved by requiring that $\EuScript{M}^{ab}(A^h)>0$. For a general classical field $A^h$ we can use Wick's theorem to invert the operator $\EuScript{M}^{ab}(A^h)$. In momentum space, one finds \cite{Capri:2015ixa,Capri:2012wx}
\begin{equation}\label{nop0}
  \Braket{p | \frac{1}{\EuScript{M}^{ab}(A^h)} |p}=\mathcal{G}^{ab}(A^h, p^2)=\frac{\delta^{ab}}{N^2-1}\mathcal{G}^{cc}(A^h,p^2)=\frac{\delta^{ab}}{N^2-1}\frac{1+\sigma(A^h,p^2)}{p^2}\;.
\end{equation}
At zero momentum, we find consequently \cite{Capri:2012wx}
\begin{eqnarray}\label{nop}
\sigma(A^h,0)&=&-\frac{g^2}{VD(N^2-1)}\int \frac{\d^D k}{(2\pi)^D}\frac{d^D q}{(2\pi)^D} A_\mu^{h,ab}(-k) \left[(\EuScript{M}(A^h))^{-1}\right]^{bc}_{k-q}A_\mu^{h,ca}(q)\;,\nonumber
\end{eqnarray}
and this leads to the exact identification

\begin{equation}
\sigma(A^h,0)=\frac{H(A^h)}{VD(N^2-1)}\;.
\end{equation}

\noindent At the level of expectation values, we can rewrite eq.~\eqref{nop0} as
\begin{eqnarray}\label{nop2}
\mathcal{G}^h(p^2)= \braket{ \mathcal{G}^{aa}(A^h, p^2)}^{conn}=\frac{1}{p^2(1-\braket{\sigma(A^h,p^2)}^{1PI})},
\end{eqnarray}
so that we must impose at the level of the path integral
\begin{equation}\label{nopole}
\sigma(0)\equiv\braket{\sigma(A^h,0)}^{1PI}< 1
\end{equation}
to ensure a positive operator\footnote{We emphasize here again, see also Sect.~III, that the removal of zero-modes of $\EuScript{M}(A^h)$ implies the elimination of a large class of zero modes of the operator $\EuScript{M}(A)$. The underlying argumentation can be found in \cite{Capri:2015ixa}. To avoid confusion, this also means that the quantity $\sigma(k)$ introduced in eq.~\eqref{pool1} is not referring to the (inverse) Faddeev-Popov ghost propagator for general $\alpha$. The connection with the ghost self-energy is only valid for the Landau gauge $\alpha=0$.} $\EuScript{M}(A^h)$.

\begin{figure}[t]\label{ghost}
    \begin{center}
        \scalebox{0.5}{\includegraphics{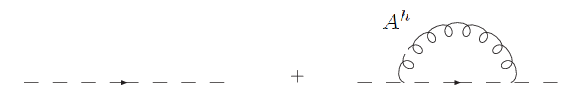}}
        \caption{The leading order correction to $\EuScript{M}^{-1}(A^h)$. The wiggled line represents a $\braket{A^h A^h}$ propagator, the broken line represents the tree level approximation to $\EuScript{M}^{-1}(A^h)$, viz.~$\frac{1}{p^2}$ in momentum space. }
    \end{center}
\end{figure}

We will now show that in the presence of the extra mass scale $M^2$, it is impossible to comply with the necessary condition \eqref{nopole} in $D=2$.
It is sufficient to work at leading order, as the problem will already reveal itself at this order. Since this corresponds to working at order $g^2$ with two factors of $g$ already coming from the term $gf^{abc}A^{h,c}_\mu\p_\mu$ in the operator $\EuScript{M}^{ab}(A^h)$, we may cut off the expansion of $A^h$ at order $g^0$, i.e.~use the approximation \eqref{lcg3}. Doing so, we find at leading order (see also Fig.~1)
\begin{eqnarray}\label{pool1}
\sigma(k) &=& g^2N \frac{k_{\mu} k_{\nu}}{k^2} \int \frac{\d^2
q}{(2\pi)^2} \frac{1}{(k-q)^2} \frac{q^2 + M^2}{q^4 + (M^2+m^2)q^2 +
\lambda^4}\left(\delta_{\mu\nu}-\frac{q_\mu q_\nu}{q^2}\right)\;.
\end{eqnarray}
We set here $\lambda^4=2g^2N \gamma^4+m^2M^2$.  The quantity $\frac{q^2 + M^2}{q^4 + (M^2+m^2)q^2 +
\lambda^4}$ is the transversal piece of the would-be Refined Gribov-Zwanziger gluon propagator in $D=2$, see  also \eqref{prop2}.

The above integral $\sigma(k)$ can be evaluated exactly quite easily by using polar coordinates. Choosing the $q_x$-axis along
$\vec{k}$, we get
\begin{eqnarray}\label{pool2}
\sigma(k) &=& \frac{g^2N}{4\pi^2}\int_0^{\infty}q\d
q\frac{q^2+M^2}{q^4+(M^2+m^2)q^2+\lambda^4}\int_{0}^{2\pi}\d\phi
\frac{1}{k^2+q^2-2qk\cos\phi}(1-\cos^2\phi)\nonumber\\
&=&\frac{g^2N}{4\pi}\left(\frac{1}{k^2}\int_0^{k}\frac{q(q^2+M^2)}{q^4+(M^2+m^2)q^2+\lambda^4}\d
q+\int_k^{\infty}\frac{q^2+M^2}{q(q^4+(M^2+m^2)q^2+\lambda^4)}\d
q\right)\;,
\end{eqnarray}
where we employed $\vec{k}\cdot\vec{q}=kq\cos\phi$, next to the integral
\begin{equation}
    \int_{0}^{2\pi}\d\phi
\frac{1-\cos^2\phi}{k^2+q^2-2qk\cos\phi}=  \frac{\pi}{q^ 2}\theta(q^2-k^2)+\frac{\pi}{k^ 2}\theta(k^2-q^2)\;.
\end{equation}
From the integrals appearing in \eqref{pool2}, we can extract the leading small $k^2$ behavior to be
\begin{equation}\label{pool3}
    \left.\sigma(k)\right|_{k^2\sim 0}\sim
    -\frac{g^2N}{8\pi}\frac{M^2}{\lambda^4}\ln(k^2)\;.
\end{equation}
Since $M^2\geq 0$, we unequivocally find that $\sigma(k^2)$ will become (much) larger than $1$ if the momentum gets too small for $M^2>0$, that is it would become impossible to
fulfill condition \eqref{nopole} and thus to ensure the positivity of $\EuScript{M}(A^h)$.

We are thus forced to conclude that $M^2=0$. Notice however that we are not able to prove that $m^2=0$. Indeed, if $M^2=0$, we are already back to the scaling case irrespective of the value for $m^2$. Scaling implies a vanishing of the transversal gluon form factor at zero momentum, which is in general sufficient to eliminate IR problems in the ghost form factor $\sigma(k^2)$, see \cite{Cucchieri:2012cb} for a general discussion.  Only an explicit discussion of the effective potential of the condensate related to $m^2$ (that is, $\braket{A^h A^h}$) will reveal whether it can be introduced or not into the theory. Though, this will not affect the conclusion that in $D=2$, a masssive/decoupling behavior is excluded.

\section{The gluon propagator} \label{Gprop}
As discussed in Sect.~\ref{RGZ}, the restriction of the path integral to a suitable region which is free of a large set of Gribov copies and is intimately related to the introduction of the Gribov parameter $\gamma$ generates dynamically dimension-two condensates. This generation is consistent in $D=3,4$, while in $D=2$ is absent. In this way, the gluon propagator is further affected by the introduction of such operators.

In $D=3,4$, the tree-level gluon two-point function is\footnote{In this expression we should keep in mind the meaning of indices and dimensions for different choices of $D$}
\begin{equation}
\langle A^{a}_{\mu}(k)A^{b}_{\nu}(-k)\rangle_{D=3,4} = \delta^{ab}\left[\frac{k^2+{M}^2}{(k^2+{m}^2)(k^2+{M}^2)+2g^2\gamma^4N}\left(\delta_{\mu\nu}-\frac{k_{\mu}k_{\nu}}{k^2}\right)+\frac{\alpha}{k^2}\frac{k_{\mu}k_{\nu}}{k^2}\right]\,.
\label{prop1}
\end{equation}
and in $D=2$,
\begin{equation}
\langle A^{a}_{\mu}(k)A^{b}_{\nu}(-k)\rangle_{D=2} = \delta^{ab}\left[\frac{k^2}{k^4+2g^2\gamma^4N}\left(\delta_{\mu\nu}-\frac{k_{\mu}k_{\nu}}{k^2}\right)+\frac{\alpha}{k^2}\frac{k_{\mu}k_{\nu}}{k^2}\right]\,.
\label{prop2}
\end{equation}
As is clear from (\ref{prop1}) and (\ref{prop2}), the longitudinal part of the tree level gluon propagator is not affected by non-perturbative effects, \textit{i.e.}, it has the same form as in the standard Faddeev-Popov quantization scheme. It is ensured by the non-perturbative BRST symmetry to hold to all orders and, therefore, is not a peculiarity of the tree-level approximation. A rigourous proof based on the corresponding Ward identities will be presented elsewhere \cite{Capri}, but we can already provide a path integral proof. We add a source term $\int \d^dx J^a b^{h,a}$ to the action  $S^{\mathrm{LCG}}_{\mathrm{RGZ}}$ to write (suppressing color indices)
\begin{equation}\label{6dd}
\braket{b^h(x) b^h(y)}=\left.\frac{\delta^2}{\delta J(y) \delta J(x)}\int [\EuScript{D}\varphi][\EuScript{D} b^h]e^{-S^{\mathrm{LCG}}_{\mathrm{RGZ}}}\right\vert_{J=0}\,.
\end{equation}
As the $b^h$-field appears at most quadratically, we find exactly
\begin{equation}\label{5dd}
     \int [\EuScript{D}\textrm{fields}]e^{-S}= \int [\EuScript{D}\textrm{fields}]e^{-\int \d^Dx\left(\frac{1}{2\alpha} (\p A)^2+\frac{1}{\alpha}J\p A+\frac{J^2}{2\alpha}+\textrm{rest}\right)}\,.
\end{equation}
or, using \eqref{6dd}
\begin{equation}\label{7dd}
   \braket{b^h(x) b^h(y)}=\frac{1}{\alpha^2}\braket{\p A(x) \p A(y)}-\frac{\delta(x-y)}{\alpha}\,.
\end{equation}
Since we also have
\begin{equation}\label{8dd}
   \braket{b^h(x) b^h(y)}=\braket{s_{\gamma^2}(\bar c(x) b^h(y) )}=0
\end{equation}
because of the non-perturbative BRST symmetry generated by $s_{\gamma^2}$, combination of \eqref{7dd} and \eqref{8dd} necessarily gives
\begin{equation}\label{7ddbis}
   0=\frac{1}{\alpha^2}\braket{\p A(x) \p A(y)}-\frac{\delta(x-y)}{\alpha}\,.
\end{equation}
which becomes in momentum space
\begin{equation}\label{8}
\langle A^{a}_{\mu}(k)A^{b}_{\nu}(-k)\rangle =D_T(k^2)\left(\delta_{\mu\nu}-\frac{k_\mu k_\nu}{k^2}\right)+\alpha\frac{k_\mu k_\nu}{k^4}
\end{equation}
where the non-trivial information is encoded in the transverse form factor $D_T(k^2)$

For the transverse component of the gluon propagator, we see that a \textit{decoupling} like behavior for $D=3,4$ is apparent, \textit{i.e.}, it has a non-vanishing form factor for zero momentum, while in $D=2$, a \textit{scaling} like behavior is observed.

This result is completely analogous to what happens in the Landau gauge. From (\ref{prop1}) and (\ref{prop2}), since the transverse part does not depend of $\alpha$, it has to be equal to the Landau gauge result, a particular choice $\alpha=0$. In this framework, $D=3,4$, the Gribov-Zwanziger action in linear covariant gauges refines and a decoupling behavior for the gluon propagator is obtained. On the other hand, the usual Gribov-Zwanziger action seems to describe the gluon propagator behavior in $D=2$, which is scaling like.

\section{A short look at the ghost propagator} \label{ghost-0}

Having worked out the expression of the gluon propagator in $D=4$, eq.\eqref{prop1}, we can have a short preliminary look at the ghost propagator. We limit here ourselves to the one-loop order, leaving a more complete and exhaustive analysis for future investigation.

For the one-loop ghost propagator in $D=4$, we have
\begin{equation}
\frac{1}{N^2-1} \sum_{ab} \delta^{ab}\langle {\bar c}^a(k) c^b(-k) \rangle_{1-loop} = \frac{1}{k^2} \frac{1}{1 - \omega(k^2)}  \;, \label{ghost-1}
\end{equation}
where
\begin{equation}
\omega(k^2) = \frac{Ng^2}{k^2(N^2-1)} \int \frac{d^4q}{(2\pi)^4} \frac{k_\mu (k-q)_\nu}{(k-q)^2} \langle A^{a}_{\mu}(q)A^{a}_{\nu}(-q)\rangle \;. \label{ghost-2}
\end{equation}
From expression \eqref{prop1}, we get
\begin{equation}
\omega(k^2) = \omega^T(k^2) + \omega^L(k^2)  \;, \label{ghost-3}
\end{equation}
where $\omega^T(k^2) $ stands for the contribution corresponding to the transverse component of the gluon propagator, {\it i.e.}
\begin{equation}
\omega^T(k^2) = {Ng^2}\frac{k_\mu k_\nu}{k^2}  \int \frac{d^4q}{(2\pi)^4} \frac{1}{(k-q)^2} \frac{q^2+{M}^2}{(q^2+{m}^2)(q^2+{M}^2)+2g^2\gamma^4N}\left(\delta_{\mu\nu}-\frac{q_{\mu}q_{\nu}}{q^2}\right) \;, \label{ghost-4}
\end{equation}
while $\omega^L(k^2)$ is the contribution stemming from the longitudinal component, namely
\begin{equation}
\omega^L(k^2) = \alpha \frac{Ng^2}{k^2} \int \frac{d^4q}{(2\pi)^4} \frac{k_\mu (k-q)_\nu}{(k-q)^2} \frac{q_\mu q_\nu}{q^4}   \;. \label{ghost-5}
\end{equation}
Employing dimensional regularization in the $\MSbar$ scheme, expression \eqref{ghost-5} yields
\begin{equation}
\omega^L(k^2) = \alpha \frac{Ng^2}{64 \pi^2} \log{\frac{k^2}{{\bar \mu}^2} }\;. \label{ghost-6}
\end{equation}
This result for $\omega^L(k^2)$ obviously coincides with the standard perturbative result at one loop. It is worth underlining that the result \eqref{ghost-6} is a consequence of the non-trivial fact that the longitudinal component of the gluon propagator is left unmodified by the addition of the horizon function $H(A^h)$, eq.~\eqref{lcg1}. The presence of terms of the type of eq.\eqref{ghost-6} seems therefore unavoidable when evaluating the ghost form factor for non-vanishing values of the gauge parameter $\alpha$. When passing from the $1PI$ Green function to the connected one, such terms should lead to a ghost form factor which is severely suppressed in the infrared region $k^2\sim 0$ with respect to the case of the Landau gauge, {\it i.e.} $\alpha=0$, as discussed recently within the framework of the Dyson-Schwinger equations \cite{Huber:2015ria,Aguilar:2015nqa}. We hope to report soon on this relevant issue. 

An important issue that deserves further study when we will attempt to compare with (no yet available) lattice data \footnote{We thank A.~Cucchieri, D.~Binosi, O.~Oliveira, and T.~Mendes for a discussion about this.} for the ghost propagator in linear covariant gauges concerns the precise definition of what is meant with ghost propagator if $\alpha\neq0$ (i.e.~outside of the Landau gauge). As $\EuScript{M}^{ab}(A)$ is not a Hermitian operator when $A$ is not transverse, neither will its inverse be, viz.~the ghost propagator defined via eq.~\eqref{ghost-1}. This has its consequences on how to define the Hermitian conjugate of the Faddeev-Popov ghost and antighost, to allow for a Hermitian formulation of the gauge action in the linear covariant gauge. For $\alpha\neq0$, the Faddeev-Popov ghost $c$ and anti-ghost $\bar c$ are to be chosen (in our conventions) to be independent and real, resp.~purely imaginary\footnote{A nice and far more detailed discussion of these matters can be found in \cite{Alkofer:2000wg}.}. The operator coupled to it, cf.~the action of expression \eqref{landau7}, is then effectively a matrix operator $\left(
                                                                                                     \begin{array}{cc}
                                                                                                       0 & D\p  \\
                                                                                                       -\p D & 0 \\
                                                                                                     \end{array}
                                                                                                   \right)$, which can be used to introduce a Hermitian ghost $1PI$ matrix and ensuing ghost propagator matrix. To our understanding, given that the lattice ghost propagator is computed via the eigenvalues of the (supposedly Hermitian) Faddeev-Popov matrix, we expect both analytical and lattice formulation of the ghost propagator should be focusing on evaluating the matrix ghost propagator, at least when we wish to compare with (future) lattice data.

\section{Conclusions}
In this paper, we extended the result of \cite{Capri:2015ixa} to take into account the presence of dimension-two condensates in the non-perturbative BRST-invariant Gribov-Zwanziger action in linear covariant gauges. As discussed, these condensates arise consistently in $D=3,4$, and a Refined Gribov-Zwanziger action which is also invariant under suitable non-perturbative BRST transformations was presented. In $D=2$, the computation of the one-loop order vacuum energy alerts that it seems impossible to introduce such vacuum condensates due to IR singularities. We made this argument more strict by showing that, again due to $D=2$ IR singularities, it is impossible to fulfill the premise underlying the whole Gribov-Zwanziger construction, {\it i.e.} positivity of the  Faddeev-Popov operator of the gauge-invariant non-local field $A^h$.  This phenomenon is completely analogous to what happens in the Landau, maximal Abelian and Coulomb gauges, in the Gribov-Zwanziger framework, \cite{Dudal:2008rm,Dudal:2008xd,Guimaraes:2015bra,Capri:2015pfa}. Such results seem to indicate that this behavior is much more universal than just a coincidence. On the other hand, it is no surprise that the situation in $D=2$ is manifestly different. Infrared divergences are known to spoil results that generally apply in $D>2$, let one only think about the Coleman-Mermin-Wagner theorem where infrared divergent integrals play a key role as well.

The introduction of condensates in $D=3,4$ modifies the transverse part of the gluon propagator. In particular, a massive (decoupling) behavior is observed, namely, a finite non-vanishing value for its form factor at zero momentum. This result is in agreement with the most recent lattice data of linear covariant gauges, \cite{Cucchieri:2009kk,Bicudo:2015rma}, which are limited to $D=4$ so far. In $D=2$, however, the transverse part only receives non-perturbative corrections from the restriction of the path integral domain, being free of the introduction of condensates. Therefore, the non-perturbative BRST-invariant Gribov-Zwanziger framework for the linear covariant gauges predicts in dimension $D=2$ a Gribov-like gluon propagator, namely: a scaling behavior, with a vanishing value at zero momentum.

It is instructive to reconsider here the origin of the difference between scaling and massive/decoupling solution. This is solely based on having $M^2\neq0$, or more precisely, of having a mass term of the type $M^2 (\bar\phi_\mu^{ab} \phi_\mu^{ab}-\bar\omega_\mu^{ab}\omega_{\mu}^{ab})$. One could wonder if this is the only possibility. The answer to this query is definitely no. However, qualitatively the answer is always the same, there is decoupling whenever a mass term in the $(\bar\phi,\phi)$-sector is introduced (dynamically). Two more general options were explored in the literature so far. In \cite{Dudal:2011gd}, different masses were allowed to couple to $\bar\phi_\mu^{ab}\bar\phi_\mu^{ab}$, $\phi_\mu^{ab}\phi_\mu^{ab}$ and $\bar\phi_\mu^{ab}\phi_\mu^{ab}$. In \cite{Gracey:2010cg}, different contractions between the color indices carried by $\bar\phi_\mu^{ab}$, $\phi_\mu^{ab}$, $\bar\omega_\mu^{ab}$ and $\omega_\mu^{ab}$ were allowed, making use of the available $SU(N)$ color tensors. In both cases, it was reported that the gluon propagator displays a massive behavior. We can however never exclude the possibility of a kind of delicate fine-tuning of several allowed ``refinement'' mass scales, which would give a scaling solution after all when all orders are considered. Such scenario has not been found till now.

Since an increasing attention is being devoted to linear covariant gauges in recent years due to increasing insights in both, we believe such results are very important not only for the Gribov-Zwanziger point of view, but also for other approaches as lattice and functional methods. The interplay between these frameworks has shown to be very productive in the Landau gauge and the situation should not be different for the linear covariant gauges.

At last, it would also be interesting to revisit the explicit determination, via appropriate gap equations, of the (now gauge-invariant) $D=2$ condensates\footnote{Indeed, $m^2$ corresponds to $\braket{A^h A^h}$ which is gauge invariant by construction, $M^2$ to $\braket{\bar\phi\phi-\bar\omega\omega}$ which is now also $s_{\gamma^2}$-closed but not exact, and the Gribov parameter $\gamma^4$ itself is also gauge invariant by virtue of the horizon condition \eqref{lcg4}.}. We hope to come back to this issue in future work. In particular would a gauge-invariant $D=2$ gluon condensate be of considerable phenomenological interest \cite{Gubarev:2000eu,Gubarev:2000nz,Chetyrkin:1998yr,Boucaud:2001st,Verschelde:2001ia}.

\section*{Acknowledgments}

The Conselho Nacional de Desenvolvimento Cient\'{i}fico e Tecnol\'{o}gico\footnote{ RFS is a level PQ-2 researcher under the program \emph{Produtividade em Pesquisa}, 308845/2012-9. MSG is a level PQ-2 researcher under the program \emph{Produtividade em Pesquisa}, 307905/2014-4. LFP is a BJT fellow from the Brazilian program Ci\^encia sem Fronteiras (Grant No. 301111/2014-6).} (CNPq-Brazil), The Coordena\c c\~ao de Aperfei\c coamento de Pessoal de N\'ivel Superior (CAPES) and the Pr\'o-Reitoria de Pesquisa, P\'os-Gradua\c c\~ao e Inova\c c\~ao (PROPPI-UFF) are acknowledged for financial support. ADP expresses his gratitude to the TPP division of SISSA and to the Perimeter Institute for Theoretical Physics for hospitality and support. DD and ADP are also thankful to the CPT of \'Ecole Polytechnique for hospitality during the final stages of this work.  This research was supported in part by Perimeter Institute for Theoretical
Physics. Research at Perimeter Institute is supported by the Government of Canada through Industry Canada and by the Province of Ontario through the Ministry of Research and Innovation.


\end{document}